\documentclass[12pt,preprint,longabstract]{aastex}
\usepackage{lscape}

\def\beq{\begin{equation}}
\def\enq{\end{equation}}
\def\bea{\begin{array}}
\def\ena{\end{array}}

\begin{document}

\title{The Mass of the Black Hole in Cygnus X-1}

\author{Jerome A. Orosz}
\affil{Department of Astronomy, San Diego State University,
5500 Campanile Drive, San Diego, CA 92182-1221}
\email{orosz@sciences.sdsu.edu}

\author{Jeffrey E. McClintock}
\affil{Harvard-Smithsonian Center for Astrophysics, 60 Garden Street,
Cambridge, MA 02138}
\email{jem@cfa.harvard.edu}

\author{Jason P. Aufdenberg}
\affil{Physical Sciences Department, Embry-Riddle Aeronautical University,
600 S. Clyde Morris Blvd., Daytona Beach, FL 32114}
\email{aufded93@erau.edu}

\author{Ronald A. Remillard}
\affil{Kavli Institute for Astrophysics and Space Research, 
Massachusetts Institute of Technology,
Cambridge, MA 02139-4307}
\email{rr@space.mit.edu}

\author{Mark J. Reid, Ramesh Narayan, Lijun Gou}
\affil{Harvard-Smithsonian Center for Astrophysics, 60 Garden Street,
Cambridge, MA 02138}
\email{reid@cfa.harvard.edu,narayan@cfa.harvard.edu,lgou@cfa.harvard.edu}

\begin{abstract}
Cygnus X-1 is a binary star system that is comprised of a black hole
and a massive giant companion star in a tight orbit.  Building on our
accurate distance measurement reported in the preceding paper, we
first determine the radius of the companion star, thereby constraining
the scale of the binary system.  
To obtain a full dynamical model of the binary, we use an
extensive collection of optical photometric and spectroscopic
data taken from the literature.  By using all of the available observational
constraints,  
we show that
the orbit is slightly eccentric
(both the radial velocity and photometric data independently confirm this
result) and that the companion star rotates roughly 1.4 times its
pseudosynchronous value.
We find a black hole mass of
$M =14.8\pm 1.0\,M_{\odot}$, a
companion mass of
${M}_{\rm opt}=19.2\pm 1.9\,M_{\odot}$,  
and the angle of inclination of
the orbital plane to our line of sight of $i=27.1\pm0.8$~deg.
\end{abstract}

\section{Introduction}

During the past 39 years, many varied estimates have been made of the
mass $M $ of the black hole in Cygnus X-1.  At one extreme, acting as a
devil's advocate against black hole models, 
\cite{tri+1973} proposed a model 
based on a distance $D\sim1$ kpc which gave
a low mass of $M  \lesssim 1\,M_{\odot}$, 
suggestive of a neutron star or white
dwarf, not a black hole.
Several other low-mass models
are summarized, considered, and found wanting by 
\cite{bol+1975}.  By contrast, all conventional binary models that
assume the secondary companion is a massive O-type
supergiant find a large---but uncertain---mass for the 
compact object that significantly exceeds 
the maximum stable mass
for a neutron star of $\approx 3\,M_{\odot}$ 
\citep{kal+1996},
hence requiring a black hole.  
For
example, using geometrical arguments, 
\citet{pac+1974} computed the minimum mass
for the compact object as a function of the distance and found
$M>3.6\,M_{\odot}$ for $D>1.4$ kpc.
Based on dynamical modeling \cite{gie+1986}
found $M>7\,M_{\odot}$ and a probable mass of 
$M_{\rm opt}=16\,M_{\odot}$ for
the companion star, and \cite{nin+1987} found
$M =10\pm1~\,M_{\odot}$ (by assuming 
$M_{\rm opt}=20\,M_{\odot}$).

However, these mass estimates, and all such estimates that have been
made to date, are very uncertain because they are based on
unsatisfactory estimates of the distance to Cygnus 
X-1 \citep{rei+2010}.  The strong effect of distance on the model
parameters is obvious from an inspection of Table~4 in
\cite{cab+2009} 
\cite[and also Table 1 of][]{pac+1974}.  
For the dynamical model favored by \citet{cab+2009}, 
and over the wide range of
distances they consider, 1.1--2.5~kpc, the radius 
of the companion star and the mass of the black
hole are seen to vary by factors of 2.3 and 11.7, respectively.
Thus, in order to obtain useful constraints on the system parameters,
it is essential to have an accurate value of the source distance, as
we have demonstrated for two extragalactic black hole systems that
contain O-type supergiants, M33 X-7 \citep{oro+2007} and LMC X-1
\citep{oro+2009}, whose distances are known to 
several percent  accuracy
via the cosmic distance ladder.

In this paper, we use a distance from a trigonometric
parallax measurement for Cygnus X-1 
\citep{rei+2010},
which is accurate to $\pm6$\%, and previously-published optical data
to build a complete dynamical model of the Cygnus X-1 binary system.
Not only are we able to strongly constrain the principal parameters of
the system, we are also able to obtain the first constraints on the
orbital eccentricity, $e$, and 
the deviation of the period of rotation of the
companion star from the orbital period, characterized by
the non-synchronous rotation parameter,
$\Omega$ \citep{oro+2009}.  Of principal interest are our precise
determinations of the black hole mass $M$ and the orbital inclination
angle $i$.  As we show in the paper that follows
\citep{gou+2010}, it
is our accurate values for the three parameters 
$M $, $i$ and $D$ that
are the key to determining the spin of the black hole.

\section{Dynamical Modeling}

The mass of the black hole can be easily determined once we know both
its distance from the center of mass of the O-star and the orbital
velocity of the star.  Since optical spectroscopy only gives us the
radial component of velocity, we must also determine the inclination
of the orbital plane relative to our line of sight in order to infer
the orbital velocity.  Furthermore, since the star orbits the center
of mass of the system, we must also obtain the separation between the
two components.  We determine the needed quantities using our
comprehensive modeling code
\citep{oro+2000}.  Our model, which is
underpinned by our new measurement of the distance, makes use of all
relevant observational constraints in a self-consistent manner.  
We discuss details of the data and modeling below.

\subsection{Optical Data Selection}

There is no shortage of published observational data for Cygnus X-1 in
the literature.  \cite{bro+1999a} provide $U$, $B$,
and $V$ light curves containing nearly 27 years worth of observations
(1971-1997) from the Crimean Laboratory of the Sternberg Astronomical
Institute.  The binned light curves contain 20 points each and are
phased on the following ephemeris:
\begin{equation}
{\rm Min ~I} =
{\rm JD~}2441163.529(\pm0.009)
+
5.599829(\pm0.000016)E,
\end{equation}
where Min~I is the time of the inferior conjunction of the
O-star, and $E$ is the cycle number.
In addition, 
\cite{bro+1999a} provide 421 radial
velocity measurements, which are
phased to the same ephemeris.  The light and
velocity curves were kindly sent to us by C. Brocksopp.


In addition to the velocity data of Broscksopp et al., we also made
use of the radial velocities published by \cite{gie+2003}.  We fitted
a sine curve to their data and those of Brocksopp et al.\ (1999a).
Fixing the uncertainties on the individual velocity measurements to
7.47 km s$^{-1}$ and 5.06 km s$^{-1}$, respectively (these values give
$\chi^2\approx N$), we found $K=74.46\pm 0.51$ km s$^{-1}$ for the
\citeauthor{bro+1999a} data and $K=75.57\pm0.70$ km s$^{-1}$ for the
\citeauthor{gie+2003} data.  The $1\sigma$ intervals of the respective
$K$-velocities overlap, the residuals of the fits show no obvious
structure, and there is no notable difference between the residuals
of the two data sets, apart from the slightly greater scatter in the
\citeauthor{bro+1999a} data.  We therefore combined the two sets of
radial velocities while removing the respective systemic velocities
from the individual sine curves.  We phased the \citeauthor{gie+2003}
data on the above ephemeris before merging these data with the
velocity data of Broscksopp et al.  The combined data set has 529
points.

All of these velocity and light-curve data are discussed further and
analyzed in Section~2.4, where they are shown folded on the orbital
period and seen to exhibit minimal scatter about the model fits.  This
small scatter in these data sets, each spanning a few decades, attests
to the strongly dominant orbital component of variability.  Meanwhile,
Cygnus~X-1 is well-known to be variable in the radio and X-ray bands,
including major transitions between hard and soft X-ray states (see
Figure~1 in Gou et al. 2011).  This raises the question of whether any
non-orbital variability in the light-curve data could significantly
affect the component masses and other parameters determined by our
model.  (We focus on the light-curve data, which are more susceptible
to being affected by variability.)

We believe that our results are robust to such non-orbital variability
for several reasons, including the following: (1) While
\citet{bro+1999b} note that the $U$, $B$, and $V$ light curves were
correlated with each other, they found no correlation with the radio
and X-ray light curves, which is reasonable given that the
time-averaged bolometric X-ray luminosity is only $\lesssim0.3$\% of
the bolometric luminosity of the O-star (Section~2.3). (2) Although
\citet{bro+1999a} did not specifically discuss the photometric
variability seen in their 27 year data set, \citet{bro+1999b} do
discuss the multiwavelength variability of the source over a 2.5 year
period.  They noted that there is very little variability in the
optical light curves apart from the dominant ellipsoidal/orbital
modulations (see Section~2.4) and one weaker modulation with about
half the amplitude on a 142 day period (which is thought to be related
to the precession of the accretion disk).  (3) The light-curve and
velocity data give remarkably consistent results for the small (but
statistically very significant) measured values of eccentricity and
argument of periastron (Section~2.4.2).  (4) As highlighted above, the
small scatter in the folded light curves for a data set spanning 27
years is strong evidence against a significant component of
variability on time scales other than the orbital period.  In
summary, we conclude that non-orbital variability is unlikely to
significantly affect our results.

\subsection{Stellar Radius, Temperature, and Rotational Velocity}

In order to constrain the dynamical model, it is crucial to have a
good estimate of the radius of the companion star.  However, customary
methods of determining this radius fail because the Cygnus X-1 system
does not exhibit eclipses nor does the companion star fill its Roche
equipotential lobe.  We obtain the required estimate of the stellar
radius as we have done previously in our study of LMC X-1
\citep{oro+2009}.  The radius, which critically depends on distance,
additionally depends on the apparent magnitude of the O-type star
and interstellar
extinction, 
and also on the effective
stellar temperature and corresponding bolometric correction.
The absolute magnitude of the star is $M_{\rm
abs}=K+BC_K(T_{\rm eff},g)-(5\log D-5) - 0.11A_V$, where $K$ is the
apparent $K$-band magnitude, $BC_K$ is the bolometric correction for
the $K$-band, $D$ is the distance, and $A_V$ is the extinction in the
$V$-band.  
The luminosity and radius of the star in solar units are
$L=10^{-0.4(M_{\rm abs}-4.71)}$ and $R=\sqrt{L(5770/T_{\rm eff})^4}$,
respectively.
In computing these quantities, we
use $D = 1.86_{-0.11}^{+0.12}$ kpc 
\citep{rei+2010} and a $K$-band
apparent magnitude of $K=6.50\pm 0.02$ 
\citep{skr+2006}, which
minimizes the effects of interstellar extinction.  For the $K$-band
extinction, we adopt $E(B-V)=1.11\pm 0.03$ and $R_V=3.02\pm 0.03$
\citep[e.g.\ $A_V=3.35$,][]{cab+2009} and use the standard extinction law
\citep{car+1989}.  The bolometric corrections for the $K$-band were
computed using the OSTAR2002 grid of models with solar metallicity
\citep{lan+2003}.  We 
note that the $K$-band bolometric corrections for the solar metallicity
models and the models for half-solar metallicity
differ only by 0.02 dex (T.\ Lanz private communication), so our results
are not sensitive to the metallicity.
 
\begin{figure}
\epsscale{0.8}
\plotone{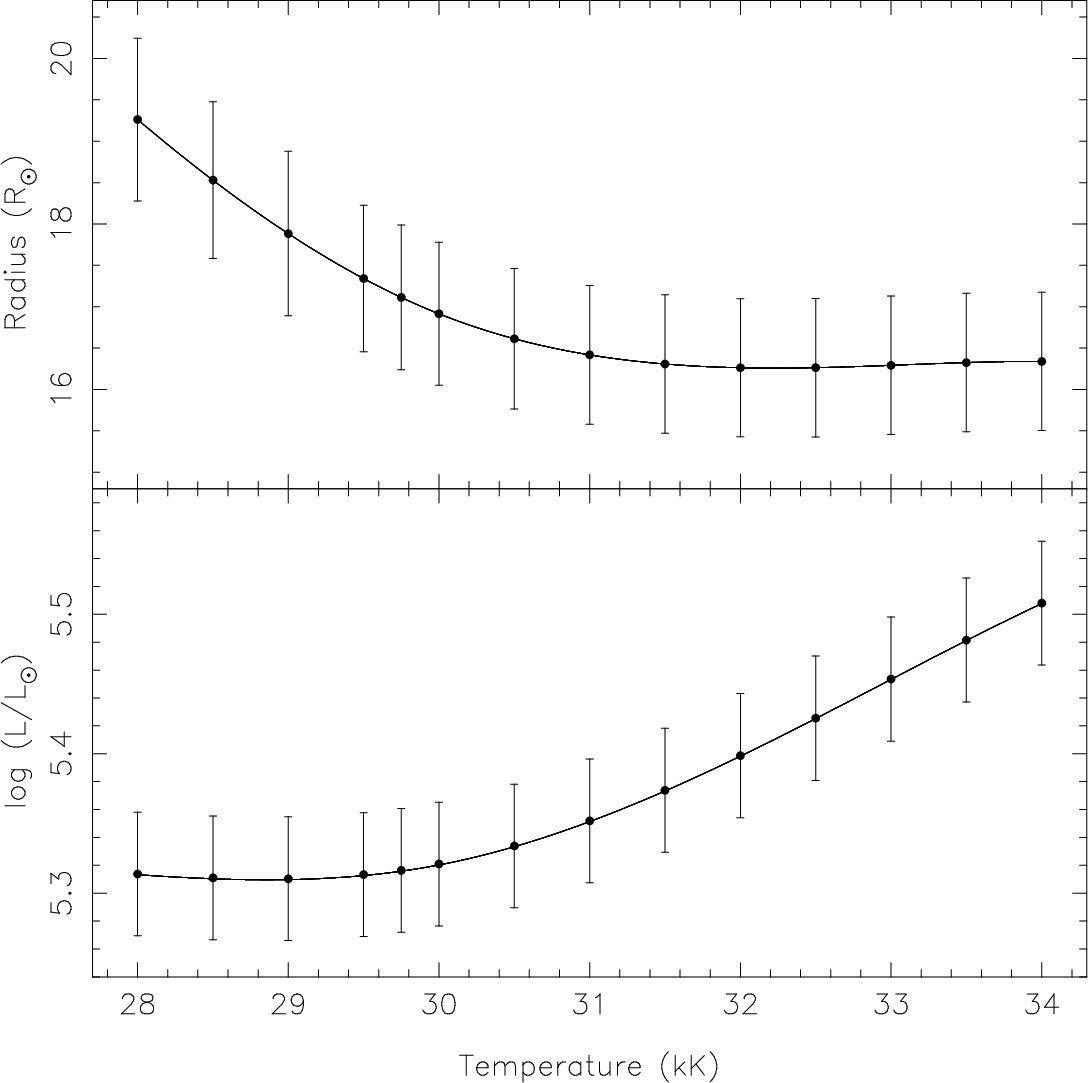}
\caption{The derived O-star radius (top) and luminosity (bottom)
as a function of the assumed
effective temperature.}
\label{plotraddist}
\end{figure}

Figure \ref{plotraddist} 
shows the derived radius and luminosity of
the star as a function of its assumed temperature in the range $28,000
\le T_{\rm eff}\le 34,000$ K.  
For $T_{\rm eff}=28,000$, the radius is $R_{\rm dist}=19.26\pm
0.98\,R_{\odot}$\footnote{We use the notation $R_{\rm dist}$ to denote the 
stellar radius derived from
the parallax distance, and $R_{\rm opt}$ to denote the stellar radius derived
from the dynamical model.}.  As the temperature increases, the
radius decreases rapidly at first, and then it plateaus midway through
the range, attaining a value of $R_{\rm dist}=16.34\pm
0.84\,R_{\odot}$ at $T_{\rm eff}=34,000$~K.  Meanwhile, the luminosity
increases with temperature, rising from $L=2.1\times 10^5\,L_{\odot}$
at $T_{\rm eff}=28,000$~K to $L=3.2\times 10^5\,L_{\odot}$ at $T_{\rm
eff}=34,000$~K.

The effective
temperature of the companion star can be determined from a detailed
analysis of UV and optical line spectra
\citep{her+1995,kar+2005,cab+2009}.
However, it is often difficult to determine a precise temperature for
O-type stars owing to a correlation between the effective temperature
$T_{\rm eff}$ and the surface gravity parameter $\log g$.  Model
atmospheres with slightly smaller values of $T_{\rm eff}$ and $\log g$
give spectra that are very similar to those obtained for slightly
higher values of these parameters.  Fortunately, $\log g$ for 
the companion 
star in Cygnus X-1
is tightly constrained for a wide range of assumptions about the
temperature, mass ratio, and other parameters because of a peculiarity
of Roche-lobe geometry
\citep{egg+1983}.  
We show below
that our dynamical model constrains the value of the
surface gravity to lie in the range $\log g=3.30-3.45$ (where $g$ is
in cm s$^{-2}$).

Based on an analysis of both optical and UV spectra, 
\cite{cab+2009} obtained for their favored model $T_{\rm
eff}=28,000\pm 2,500$~K and a surface gravity $\log g\gtrsim 3.00 \pm
0.25$, which is outside the range of values implied by our model.
Based on the plots and tables in 
\cite{cab+2009}, we estimate a best-fitting temperature of $T_{\rm
eff}=30,000\pm 2,500$~K when the surface gravity is forced to lie in
the range determined by our dynamical model.  Using optical spectra,
\cite{kar+2005} found $T_{\rm eff}=30,400 \pm 500$
K and $\log g=3.31\pm 0.07$, which is consistent with our
dynamically-determined value.  Likewise, 
\cite{her+1995} report $T_{\rm eff}=32,000$ and $\log g=3.21$ (no
uncertainties are given in their Table 1).  In the following, we adopt
a temperature range of $30,000 \le T_{\rm eff}\le 32,000$ K.

After considering several previous determinations of the projected
rotational velocity of the O-type star and corrections for
macroturbulent broadening, \cite{cab+2009}
adopt $V_{\rm rot}\sin i=95\pm 6$ km s$^{-1}$.  We use this value as a
constraint on our dynamical model, which we now discuss.

\subsection{ELC Description and Model Parameters}

The ELC (Eclipsing Light Curve)
model \citep{oro+2000} 
has parameters related to the system
geometry and parameters related to the radiative properties of the
star.  For the Cygnus X-1 models, the orbital period is fixed at
$P=5.599829$ days \citep{bro+1999a}.  
Once the values of $P$, the $K$-velocity of the O-star and its mass
$M_{\rm opt}$ are known, the scale-size of the binary (e.g., the
semimajor axis $a$) and the mass of the black hole $M$ are uniquely
determined.  With the scale of the binary set,
the radius of the star $R_{\rm
opt}$ determines the Roche-lobe filling fraction $\rho$. 
Not all values of $R_{\rm opt}$ are allowed (for a given $P$, $M_{\rm
opt}$, and $K$): If $R_{\rm opt}$ exceeds the effective radius of the
O-star's Roche lobe, we then set the value of $\rho$ to unity.

The main parameters that control the radiative properties of the
O-star are its effective temperature $T_{\rm eff}$, its gravity
darkening exponent $\beta$, and its bolometric albedo $A$.  Following
standard practice for a star with a radiative outer envelope, we set
$\beta=0.25$ and $A=1$.

The ELC model can also include optical
light from a flared accretion disk.
In the case of Cygnus X-1, the
O-star dominates the optical and UV flux \citep{cab+2009},
where the ratio of stellar flux to accretion disk flux at 5000~\AA\ is
about $10,000$.  Consequently, we do not include any optical light from
an 
accretion disk.

We turn to the question of the X-ray heating of
the supergiant star and its effect on the binary model.
The
X-ray heating is computed using the technique outlined in 
\cite{wil+1990}.  The X-ray source geometry
is assumed to be a thin disk in the
orbital plane with a radius vanishingly small compared to the
semimajor axis (this structure should not be confused with 
much larger accretion disk that potentially could 
be a source of
optical flux).  Points on the stellar surface 
``see'' the X-ray source at inclined angles, and the proper
foreshortening is accounted for.

Measurement of the broadband 
X-ray luminosity of Cygnus X-1 
($L_{\rm xbol}$; hereafter in units of $10^{37}$ erg 
s$^{-1}$,
adjusted to the revised distance of 1.86 kpc) requires
special instrumentation and  considerations.  There are soft
and hard states of Cygnus X-1 (e.g., Gou et al.\ 2011). The hard state is
especially challenging because the effective temperature of the
accretion disk is relatively low ($T <$ 0.5 keV), while the hard 
power-law component (with photon index $\sim 1.7$) must be integrated past
the cutoff energy $\sim 150$ keV (Gierlinski et al. 1997; Cadolle Bel
et al.\ 2006). Since the ground based observations (i.e., photometric
data and radial velocity measurements) are distributed over many
years, both the range and the long-term average of the X-ray
luminosity must be estimated.
 
The archive of the Rossi X-ray Timing Explorer (RXTE) contains several
thousand observations of Cygnus X-1 collected in numerous monitoring
campaigns conducted over the life of the mission.
We processed and analyzed 2343 exposure intervals (1996
January to 2011 February; mean exposure 2.2 ks) with the PCA
instrument, and we then computed normalized light curves in four
energy bands (2-18 keV) in the manner described by Remillard \&
McClintock (2006).  In the hardness-intensity diagram, the soft and
hard states of Cyg X-1 can be separated by the value of hard color
($HC$; i.e., the ratio of the normalized PCA count rates at 8.6-18.0
versus 5.0-8.6 keV), using a simple discrimination line at $HC = 0.7$.
On this basis, we determine that Cygnus X-1 is found to be in the hard
state 73\% of the time.

Zhang et al.\ (1997) studied the broadband spectra of Cygnus
X-1 with instruments of RXTE and the Compton Gamma Ray Observatory
(CGRO). During the hard and soft states of 1996, they found broad-band
X-ray luminosities in the range 1.6-2.2
for the hard state and 2.2-3.3 for the soft state.  Samples of the
hard and intermediate states during 2002-2004 with the
International Gamma-Ray Astrophysics Laboratory (INTEGRAL) yielded
$L_{\rm xbol}$ in the range 1.2-2.0 (Cadolle Bel et al.\ 2006).  Additional
measurements of the soft state with Beppo-SAX (Frontera et al.\ 2001)
found $L_{\rm xbol}$ in the range 1.7-2.1, while Gou et al.\ (2011)
analyzed bright soft state observations by ASCA and RXTE (1996) or
Chandra and RXTE (2010) to determine $L_{\rm xbol}$ in the range 3.2-4.0.

To get a rough idea how these special observations relate to
typical conditions, we used the available contemporaneous RXTE
observations to scale the measured $L_{\rm xbol}$ values against the PCA
count rates, considering hard and soft states separately.  We then
estimate that the average hard and soft states would correspond to
$L_{\rm xbol}$ values of 1.9 and 3.3, respectively. Finally, the
time-averaged luminosity would then be roughly $2.1 \times 10^{37}$
erg s$^{-1}$, which corresponds to 0.01
 of the Eddington limit
for Cygnus X-1.
This value is much smaller
than the bolometric luminosity of the O-star ($L_{\rm Bol}\approx
7.94\times10^{39}~{\rm erg~s^{-1}}$), 
and so we expect that 
X-ray heating  in Cygnus X-1 will not be 
a significant source of systematic error.
We ran some simple tests in which we increased
the bolometric X-ray luninosity
by up to an order of magnitude,
and  found the light curves to be essentially identical.

When computing the light curve, ELC integrates the various intensities
of the visible surface elements on the star.  At a given phase and
viewing angle, each surface element on the star has a temperature $T$,
a gravity $\log g$, 
and a viewing angle $\mu=\cos\theta$, where $\theta$ is
the angle between the surface normal and the line-of-sight.  ELC has a
pre-computed table of specific intensities for a grid of models in the
$T_{\rm eff}-\log g$ plane.  Consequently, no parameterized limb
darkening law is needed.
For the specific case of Cygnus X-1, the
range of temperature-gravity pairs on the star contained points that
are outside the OSTAR2002+BSTAR2006 model grids
\citep{lan+2003,lan+2007}.  We therefore computed a new grid of models,
assuming solar metallicity. 
The range in effective temperature is from 12,000 K to
40,000 K in steps of 1000 K, and the
range in the logarithm of the surface
gravity $\log g$ (cm s$^{-2}$) is from 2.00 to 3.75 in steps of 0.25
dex.  
The wavelength-dependent radiation fields $I_{\lambda} (\mu)$,
as a function of the cosine of
the emergent angle $\mu$, were computed from 232 spherical,
line-blanketed, LTE models using the generalized model stellar
atmosphere code PHOENIX, version 15.04.00E 
\citep{auf+1998}.   These models take into
consideration between 1,176,932 and 2,296,243 atomic lines in the
computation of the line opacity.
The radiation fields were computed at 26,639 wavelengths between 0.1
nm and 900,000 nm and at 228 angle points at the outer boundary of
each model structure.  The models include 100 depth points, an outer
pressure boundary of 10$^{-4}$ dyn cm$^{-2}$, an optical depth range
(in the continuum at 500 nm) from 10$^{-10}$ to 10$^{2}$, a
microturbulence of 2.0 km s$^{-1}$, and they
have a radius of $18\,R_{\odot}$.  (At the gravities of interest, the
models are very insensitive to the precise value of the stellar
radius.)  
A table of filter-integrated specific intensities 
for the Johnson $U$, $B$, and $V$ filters for use in ELC was
prepared in the manner described in \citet{oro+2000}.

As noted above,
the observational data we model include the $U$, $B$, and $V$ light
curves from \cite{bro+1999a}, and the radial
velocities from \cite{bro+1999a} and \cite{gie+2003},
which are
combined into one set.  We use the standard $\chi^2$ statistic
to measure the
goodness-of-fit of the models:
\begin{eqnarray}
\chi^2_{\rm data}  = & & \sum_{i=1}^{20}\left({U_i-U_{\rm mod}(\vec{a})\over 
\sigma_i}\right)^2
 + 
\sum_{i=1}^{20}\left({B_i-B_{\rm mod}(\vec{a})\over \sigma_i}\right)^2
+
\sum_{i=1}^{20}\left({V_i-V_{\rm mod}(\vec{a})\over \sigma_i}\right)^2 
\nonumber \\
& & + \sum_{i=1}^{529}\left({RV_i-RV_{\rm mod}(\vec{a})\over 
     \sigma_i}\right)^2. 
\end{eqnarray}
Here, the notation $U_i$ refers to the observed $U$ magnitude at a
given phase $i$; $\sigma_i$ is the uncertainty; and $U_{\rm
mod}(\vec{a})$ is the model magnitude at that same phase, given a set
of fitting parameters $\vec{a}$.  Similar notation is used for the $B$
and $V$ bands, and for
the radial velocities.  

Initially, we assumed a circular orbit and synchronous rotation, which
left us with five free parameters: the orbital inclination angle $i$,
the amplitude $K$ of the O-star's radial velocity curve (the
``$K$-velocity''), the mass of the star $M_{\rm opt}$, the radius of
the star $R_{\rm opt}$, and a small phase shift parameter $\phi$  
to account for slight errors in the
ephemerides.  We quickly found, upon an examination of the post-fit
residuals, that circular/synchronous models are inadequate and more
complex models are required.  In the end, we computed four classes of
models: 
(i) Model A has a circular orbit and synchronous rotation, and it uses
the five free parameters discussed above:
$\vec{a}=(i,k,M_{\rm opt},R_{\rm
opt},\phi)$. 
(ii) Model B has a circular
orbit and nonsynchronous rotation, requiring one additional free parameter
$\Omega$, which is
the ratio of the rotational frequency of
the O-star to the orbital frequency:
$\vec{a}=(i,k,M_{\rm opt},R_{\rm opt},\phi,\Omega)$.  
(iii) Model C has an eccentric
orbit
and synchronous rotation at periastron. 
For this model there are seven free parameters, namely the five parameters
for Model A plus the eccentricity $e$ and the argument of periastron
$\omega$:
$\vec{a}=(i,k,M_{\rm opt},R_{\rm
opt},\phi,e,\omega)$.  Finally, (iv) Model D incorporates the possibility of
nonsynchronous rotation and an eccentric orbit
and has a total of eight free parameters:
$\vec{a}=(i,k,M_{\rm opt},R_{\rm
opt},\phi,e,\omega,\Omega)$.

We also have additional observed constraints that do not apply to a
given orbital phase, but rather to the system as a whole.  These
include the stellar radius 
(computed from the
parallax distance $R_{\rm dist}$  for a given temperature) 
and the rotational velocity of
$V_{\rm rot}\sin i=95\pm 6$ km s$^{-1}$.  For Cygnus X-1, X-ray
eclipses are not observed, and hence there is no eclipse constraint.
As discussed in \cite{oro+2002}, the radius and velocity
constraints are imposed by adding additional terms to the value of
$\chi^2$:
\begin{equation}
\chi^2_{\rm con}=\left({R_{\rm opt}-R_{\rm dist}\over \sigma_{R_{\rm
      dist}}}\right)^2+ \left({V_{\rm rot}\sin i-95\over 6}\right)^2.
\end{equation}
Finally, for a given model, we have as our ultimate measure of the
goodness-of-fit
\begin{equation}
\chi^2_{\rm tot}=\chi^2_{\rm data}+\chi^2_{\rm con}.
\end{equation}

As with any fitting procedure, some care must be taken when
considering relative weights of various data sets.  In the case of
Cygnus X-1, the quality of the light curves is similar in all three
bands.  We therefore scaled the uncertainties by small factors in
order to get $\chi^2\approx N$ for each set separately.  
This resulted in mean uncertainties of
0.00341 mag for $U$, 0.00155 mag
for $B$, and 0.00226 for $V$.  
The uncertainties for the radial velocity measurements are given in
Section~2.1.
%

\subsection{ELC Fitting}

\subsubsection{Principal Results}

The fits to the light-curve and radial-velocity data for Model A
(left panels) and Model D (right panels) are shown in Figure
2, and
the best-fit values of the parameters are given for all four models in
Table 1.  These results are for $T_{\rm eff}=31,000$ K; similar
figures and tables for other temperatures are presented in the Appendix.
For a given temperature, the tabular data reveal a key trend: The
total $\chi^2$ of the fit rapidly decreases as one goes from Model A
to Model D, indicating that eccentricity and nonsynchronous
rotation are important elements,
and we therefore adopt
Model~D.  
For
final values of the fitted and derived parameters given in Table 1, we
take the weighted average of the values derived for each of the
temperatures in the range $30,000\le T_{\rm eff}\le 32,000$~K.  A
schematic diagram of the binary based on our best-fitting model is
shown in Figure 3.

\subsubsection{Need for Non-Zero Eccentricity}

In our dynamical model, the orbit is not quite circular and the stellar
rotation is not synchronous at periastron.  Although eccentric 
orbits based on light curve models have been considered in the
past (e.g. Hutchings 1978;  Guinan et al.\ 1979), most of the
recent work considers circular orbits and synchronous rotation
(e.g.\ Caballero-Nieves et al.\ 2009).  While the
eccentricity we find is small ($e=0.018 \pm 0.002$ for Model D at
$T_{\rm eff}=31000$ K), it is highly significant. 
Allowing the orbit to be eccentric results
in an improvement to the $\chi^2$ values of all three light curves and
the radial velocity curve as well.  
As a separate check on our results, we fitted the light curves 
and the velocity curve separately and found the best-fitting values
of $e$ and $\omega$ for each.  
For the light curves, which are derived from a homogeneous data set
containing more than 800 observations spanning 27 years, we find
$e=0.0249\pm 0.003$ and $\omega=305\pm 7^{\circ}$.  For the velocity
curve, we find $e=0.028\pm 0.006$ and $\omega=307\pm 13^{\circ}$.
In addition to the remarkable consistency of the results, 
note that the argument of periastron we find is not consistent with
either $90^{\circ}$ or $270^{\circ}$, which indicates that the
eccentricity is not an artifact of tidal distortions 
\citep{wil+1976,eat+2008}.

\subsubsection{Sensitivity of the Results on the Model Assumptions}

A glance at Table 1 shows some of the parameters, 
such as the inclination,
seem to change considerably among the four models.  These differences
can be explained in part by how much weight one places on the 
external constraints, namely the radius found from the parallax 
distance and the observed rotational velocity.  
If we assume synchronous rotation, then the observed radius of
$16.42\pm 0.84\,R_{\odot}$ (for $T_{\rm eff}=31,000$) and the
projected rotational velocity of $V_{\rm rot}\sin i=96\pm 6$ km
s$^{-1}$ give an inclination of $i=39.8\pm 3.9^{\circ}$.  We ran a
model $C^{\prime}$ that is identical to model C except 
we fixed the errors
on the external constraints to be a factor of 10 smaller (e.g.\
$16.42\pm 0.084\,R_{\odot}$ and $V_{\rm rot}\sin i=96\pm 0.6$ km
s$^{-1}$).  We found an inclination of $i=39.6\pm 0.4^{\circ}$, which
is consistent with the expected value.  However, the fits are much
worse, with $\chi^2=629.4$ compared to $\chi^2=614.8$.  Likewise, we
ran a model $A^{\prime}$ that is identical to model A except for the
use of the same hard external constraints, and we similarly found
$i=39.6\pm 0.4^{\circ}$ and $\chi^2=673.2$.  Thus, in this limiting
case where the radius and rotational velocity are forced to their
observed values, the inclination is determined by these quantities and
does not depend on the eccentricity.

In principle, the light curves should be sensitive to nonsynchronous
rotation since the potential includes a term containing  a factor of
$\Omega^2$, where $\Omega$ is the ratio of the stellar rotational
frequency to the orbital frequency \citep{oro+2000}.  
As shown in Table 1, we obtain improved fits to the light curve when
the rotation is allowed to be nonsynchronous: model B has a smaller
$\chi^2$ than model A (although both of these models are rejected
since the eccentricity is nonzero), and model D has
a
smaller $\chi^2$ than model C.  In the case of model D, note that
the values of $\chi^2$ for each of the $U$, $B$, and $V$ light
curves has decreased
relative to the values for model C. When
the rotation of the O-star is not synchronous, the mapping between the radius,
the inclination, and the observed projected rotational velocity
is of course  altered.  Thus in model D we are able to 
satisfy the constraints imposed both by the observed radius and the
observed rotational velocity while providing optimal fits to the
individual light curves.

Finally, we note that it is possible to achieve optimal fits to the
light curves for an eccentric orbit and synchronous rotation if
we ignore the constraints provided by the measured values of the radius 
and rotational velocity.  
However, without these constraints there are degeneracies
among solutions.  For example, two solutions with similar
$\chi^2$ values exist:
one solution has $i=21.4^{\circ}$, $R_2=22.98\,R_{\odot}$,
$V_{\rm rot}\sin i=98.4$ km s$^{-1}$, and $\chi^2=597.0$,
whereas another solution has $i=25.4^{\circ}$, $R_2=16.2\,R_{\odot}$,
$V_{\rm rot}\sin i=63.0$ km s$^{-1}$, and $\chi^2=596.5$.
In the former case, the derived
rotational velocity is consistent with the observed value,
but the derived stellar radius is much too large.  
In the latter case, the derived stellar radius is consistent with the
radius computed from the parallax distance; however, the derived
rotational velocity is much too small.
As mentioned above,
allowing the stellar rotation to be nonsynchronous allows us to 
satisfy all of the constraints while optimally fitting the light curves.

\section{Discussion and Summary}

We find masses of $M_{\rm opt}=19.16\pm 1.90\,M_{\odot}$ and $M
=14.81\pm 0.98\,M_{\odot}$ for the O-star and black hole,
respectively.   
These estimates are considerably more direct and robust
than previous ones, owing largely to the new parallax
distance.   This secure trigonometric distance
was used to derive a radius
for the O-star using the observed $K$-band
magnitude to minimize uncertainties due
to extinction and metallicity.
The derived O-star radius
strongly constrains the dynamical
model.  In addition, 
we have also used the observed rotational velocity
of the O-star as an
observational
constraint, and sought models that simultaneously
satisfy all of the constraints.
We have explored possible sources of systematic error
by
considering a wide range of possible
temperatures for the companion O-star and
circular and eccentric models separately.  The optical
light curves we used come from a long-term
program which used the same instrumentation and
should represent the mean state of the system
quite well.  Overall, the uncertainties in
the dynamical model are as about as small
as they can be in a noneclipsing system.

By way of comparison with previous work,
a recent summary of mass estimates is given in
\cite{cab+2009}.  
The four estimates given for the black hole mass that are based on
dynamical studies (see Section 1) are quite uncertain and generally
consistent with our result.
Other estimates, which are based on less certain methods, generally
imply a lower mass, $M\approx10\,M_{\odot}$.  For example,
\cite{abu+2005} report
a mass in the range $8.2\le M\le
12.8\,M_{\odot}$ from their analysis of optical spectral line
profiles.  \cite{sha+2007} found a mass in
the range $7.9\le M\le 9.5\,M_{\odot}$ using X-ray
spectral/timing data and a scaling relation based on the
dynamically-determined masses of three black holes.  Such methods,
while intriguing, are not well established and are prone to large
systematic errors compared to the time-tested and straightforward
approach that we have taken.

Our improved dynamical model for Cyg X-1 enables other studies of this
key black-hole binary.  Using our precise measurement of the distance
(Reid et al. 2011), we are now able to compute the stellar luminosity
as a function of the assumed temperature, which allows one to place
the star on a H-R diagram with some confidence.  We have also provided
precise values of the component masses, the eccentricity, and the
degree of nonsynchronous rotation, quantities which may be used to
test binary evolutionary models (such a study will be presented
elsewhere).  Finally, with our accurate values for the distance, the
black hole mass, and the orbital inclination angle, one can model
X-ray spectra of the source in order to measure the spin of the black
hole primary.  In our third paper in this series, Gou et al.\ (2011),
we show that Cyg X-1 is a near-extreme Kerr black hole.

\acknowledgments

We thank Catherine Brocksopp for sending us the optical light curves.
The work of JEM
was supported in part by NASA grant NNX11AD08G.  
This research has made use of NASA's Astrophysics Data System.

\clearpage

\begin{deluxetable}{cccccc}
\tabletypesize{\scriptsize}
\tablecaption{Cygnus X-1 Model Parameters}
\tablewidth{0pt}
\tablehead{
\colhead{parameter} & 
\colhead{Model A} & 
\colhead{Model B} & 
\colhead{Model C} & 
\colhead{Model D} & 
\colhead{adopted}}
\startdata
$i$ (deg) & $51.45\pm 7.46$ & $67.74\pm 2.83$ & $28.50\pm 2.21$ & 
$27.03\pm 0.41$ &   $27.06\pm 0.76$\\
$\Omega$  & 1.000           & $0.674\pm 0.043$ & 1.000 & 
$1.404\pm 0.099$ &  $1.400\pm 0.084$ \\
$e$       & ...             & ...             & $0.025\pm 0.003$ 
& $0.018\pm 0.002$ &  $0.018\pm 0.003$  \\
$\omega$ (deg) & ...        & ...             & $303.5\pm 5.1$ & $ 308.1\pm 5.5$
 & $307.6 \pm 5.3$ \\
$M_{\rm opt}$ ($M_{\odot}$) & $20.53\pm 2.00$ & $26.27\pm 2.41$ & 
$ 25.17\pm 2.54$ & $18.97\pm 2.15$ &  $19.16\pm 1.90$\\
$M $ ($M_{\odot}$) & $7.37\pm 1.19$ & $6.98\pm 0.39$ & $15.83\pm 1.80$ & 
$14.69\pm 0.70$ & $14.81\pm 0.98$ \\
$R_{\rm opt}$ ($R_{\odot}$)& $15.07\pm 1.26$ & $16.41\pm 0.72$ & 
$18.46\pm 0.77$
 & $16.09\pm 0.65$ & $16.17\pm 0.68$\\
$R_{\rm dist}$ ($R_{\odot}$)& $16.42\pm 0.84$ &$16.42\pm 0.84$ &$16.42\pm 0.84$ 
&$16.42\pm 0.84$ & $16.50 \pm 0.84$\\
$\log g$ (cm s$^{-2}$) & $3.394\pm 0.016$ & $3.427\pm 0.008$ & 
$3.306\pm 0.018$ 
& $3.302\pm 0.012$ & $3.303 \pm 0.018$\\
$V_{\rm rot}\sin i$ (km s$^{-1}$)& $106.52\pm 6.39$ & $92.57\pm 5.59$ & 
$79.91\pm 4.79$ & $92.99\pm 5.89$ & $92.99\pm 4.62$ \\
$\chi^2_U$  & 26.11 & 24.74 & 21.14 & 17.76 & ... \\
$\chi^2_B$  & 46.71 & 43.57 & 24.96 & 19.33 & ... \\
$\chi^2_V$  & 34.81 & 32.49 & 24.58 & 23.99 & ...  \\
$\chi^2_{\rm RV}$ & 554.13 & 551.57 & 531.83 & 536.32 & ... \\
$\chi^2_{\rm tot}$ & 668.03 & 652.54 & 614.769 & 597.67 & ...  \\
d.o.f.  &  584 & 583 & 582 & 581 & ... \\
\enddata
\tablecomments{The assumed stellar temperature is
$T_{\rm eff}=31,000$~K.
$R_{\rm opt}$ is the stellar radius derived from the models.
$R_{\rm dist}$ is the stellar radius derived from the measured parallax and
assumed temperature.  
$V_{\rm rot}\sin i=95\pm 6$ km s$^{-1}$
is the projected rotational velocity of the O-star
derived from the models.  
Model A assumes a circular orbit and synchronous rotation.
Model B assumes a circular orbit and nonsynchronous rotation.
Model C assumes an eccentric orbit and pseudo-synchronous rotation.
Model D assumes an eccentric orbit and nonsynchronous rotation.
The adopted parameters are the weighted averages of the values 
for model D
derived for
each temperature in the range of $30,000\le T_{\rm eff}
\le 32,000$ K.}
\end{deluxetable}

\clearpage

\begin{deluxetable}{rr}
\tablecaption{Cygnus X-1 Final Parameters\label{finalparm}}
\tablewidth{0pt}
\tablehead{
\colhead{parameter} &          
\colhead{value}         
}
\startdata
$i$ (deg)  &  $27.06 \pm 0.76$ \\
$\Omega$   &  $1.400\pm 0.084$ \\
$e$        &  $0.018\pm 0.003$ \\
$\omega$ (deg) &  $307.6\pm 5.3$ \\
$M_{\rm opt}$  ($M_{\odot}$) &  $19.16\pm 1.90$  \\
$R_{\rm opt}$  ($R_{\odot}$)  &  $16.17\pm 0.68$ \\
$\log g$ (cgs)  &  $3.303\pm 0.018$ \\
$M$ ($M_{\odot}$) & $14.81\pm 0.98$ \\
\enddata
\end{deluxetable}

\clearpage

\begin{figure}
\epsscale{0.8}
\plotone{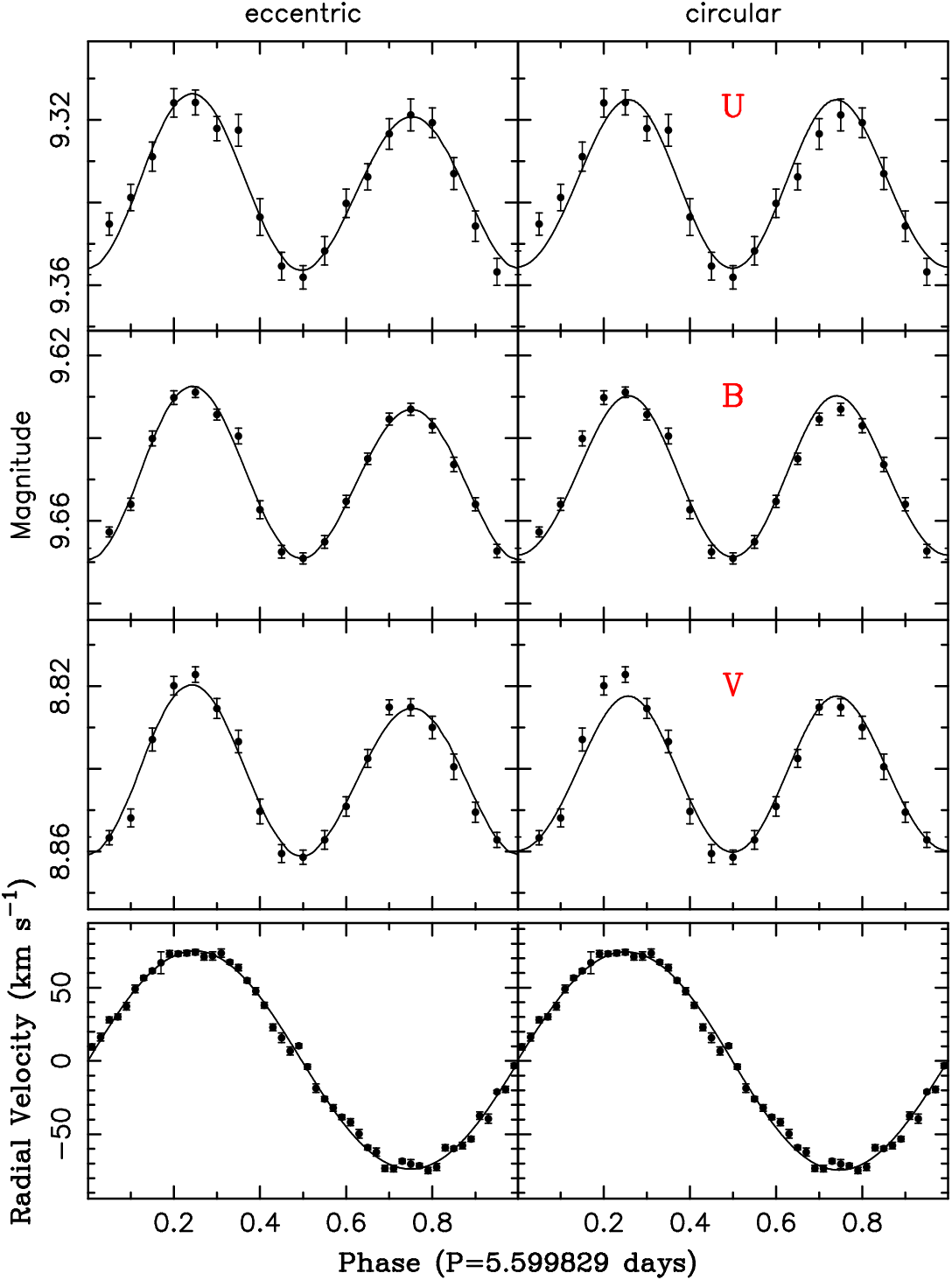}
\caption{Top: The optical light curves and
best-fitting models assuming an eccentric orbit with
$e=0.018$ (model D, left panels)
and a circular orbit (model A, right panels).  Note how much better
the unequal maxima of the light curves are accommodated by the model
that includes eccentricity.  
Bottom: The radial velocity measurements binned into 50 bins (filled
circles) and the best-fitting model for the eccentric orbit (left) and
the circular orbit (right).}
\label{fitfig}
\end{figure}

\begin{figure}
\includegraphics[scale=0.8,angle=-90]{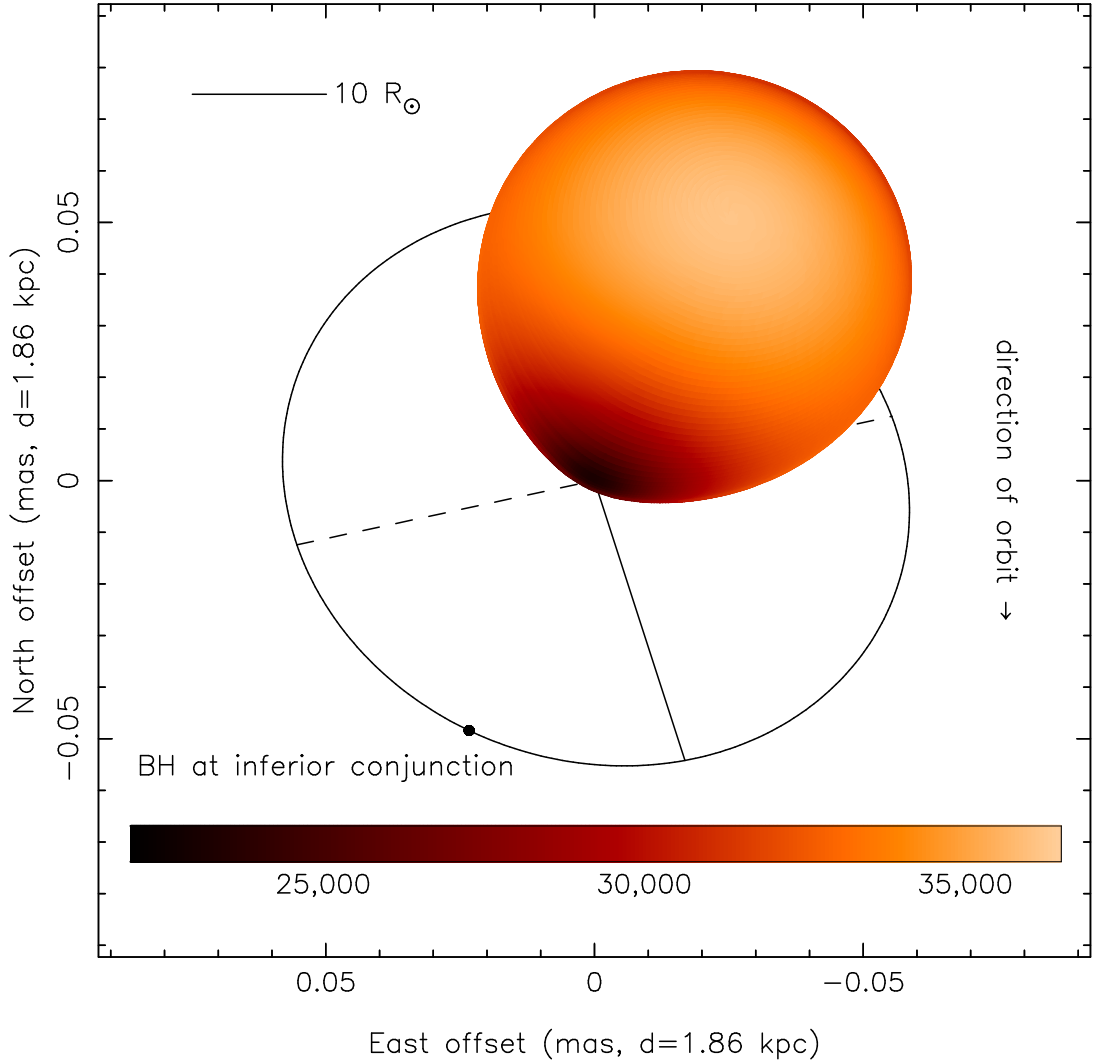}
\caption{A schematic diagram of Cygnus X-1, shown
as it would appear on the sky plane.  The offsets are in
milliarcseconds (mas), assuming a distance of 1.86 kpc.  The orbital
phase is $\phi=0.5$, which corresponds to the superior conjunction of
the O-star.  The orbit of the black hole is indicated by the ellipse,
where the major and minor axes have been drawn in for clarity (solid
line and dashed line, respectively).  The direction of the orbital
motion is clockwise, as determined by the radio observations
\protect{\citep{rei+2010}}.  The color map represents the local
effective temperature.  The star is much cooler near the inner
Lagrangian point because of the well-known effect of ``gravity
darkening'' \citep{oro+2000}.  The temperatures referred to in the main
text, Figure 1 and Table 1 refer to intensity-weighted average values.
\label{schem}}
\end{figure}

\clearpage

\appendix

\section{ELC Fitting Details}

We computed models for 14 values of $T_{\rm eff}$ between 28,000 and
34,000 K.  We fit for each temperature separately since the
derived radius from the parallax distance depends on the temperature,
and the derived radius is used as an external constraint.
For each model at each temperature, ELC's genetic code was
run twice using different initial populations and the Monte Carlo
Markov Chain (MCMC) code was run once.  The best solutions were then
refined using a simple grid search.  We computed uncertainties on the
fitted parameters and on the derived parameters (e.g.\ the black hole
mass $M$, the gravity of the O-star $\log g$, etc.) using the
procedure discussed in \cite{oro+2002}.  The results for
all 14 temperatures are shown in Table A1 and Figs.\ A1 and A2.  As
discussed in the main text, the improvement in the $\chi^2$ values as
one goes from Model A to Model D is evident.  For Model A, we
consistently find $R_{\rm opt}< R_{\rm dist}$.  Furthermore, the
rotational velocity derived from the model is consistently larger than
the observed value.  By allowing nonsynchronous rotation for the
circular orbit (as in Model B), the model-derived stellar radius
agrees with the radius computed from the distance, and the
model-derived rotational velocity agrees with the measured one.
Generally speaking, the star rotates slower than its synchronous
value.  However, an inspection of the light curves (see Figure 2 in the
main text) shows that the maximum near phase 0.25 is slightly higher
than the maximum near 0.75.  Because an ellipsoidal model predicts
maxima of equal intensity, the fit to the data is not optimal.  We
accommodate the unequal maxima by adding eccentricity to the
synchronous model (as in Model C).  However, in this case, the model
stellar radius $R_{\rm opt}$ is consistently larger than the radius
derived from the distance $R_{\rm dist}$ and the model rotational
velocity is smaller than the observed value.  By allowing
nonsynchronous rotation in the eccentric orbit (Model D), $R_{\rm
opt}$ agrees with $R_{\rm dist}$ and the model rotational velocity
agrees with the measured value.

\clearpage

\setcounter{table}{0}
\renewcommand{\thetable}{A\arabic{table}}

\pagestyle{empty} 
\begin{deluxetable}{rcccccccccccccccc}
\tabletypesize{\scriptsize}
\rotate
\tablecaption{Cygnus X-1 Model Parameters\label{parm}}
\tablewidth{0pt}
\tablehead{
\colhead{$T_{\rm eff}$} &          
\colhead{model}                  &
\colhead{$i$}                  &
\colhead{$\Omega$}                  &
\colhead{$e$}                  &
\colhead{$\omega$}                  &
\colhead{$M_{\rm opt}$}                  &
\colhead{$M$}                  &
\colhead{$R_{\rm opt}$\tablenotemark{a}}                  &
\colhead{$R_{\rm dist}$\tablenotemark{b}}                  &
\colhead{$\log g$}                  &
\colhead{$V_{\rm rot}\sin i$\tablenotemark{c}}                  &
\colhead{$\chi^2_U$}                  &
\colhead{$\chi^2_B$}                  &
\colhead{$\chi^2_V$}                  &
\colhead{$\chi^2_{\rm RV}$}                  &
\colhead{$\chi^2_{\rm tot}$}       \\
\colhead{(K)}         &
\colhead{}         &
\colhead{(deg)}         &
\colhead{}         &
\colhead{}         &
\colhead{(deg)}         &
\colhead{($M_{\odot}$)}         &
\colhead{($M_{\odot}$)}         &
\colhead{($R_{\odot}$)}         &
\colhead{($R_{\odot}$)}         &
\colhead{(cgs)}         &
\colhead{(km s$^{-1}$)}         &
\colhead{}         &
\colhead{}         &
\colhead{}         &
\colhead{}         &
\colhead{}         
}
\startdata
28000 & A\tablenotemark{d} 
                   & $48.96$
                   & $1.000$
                   & \nodata
                   & \nodata
                   & $26.61$
                   & $ 9.01$
                   & $16.88$
                   & $19.26$
                   & $3.408$
                   & $115.08$
                   & $31.13$
                   & $59.68$
                   & $41.26$
                   & $556.00$
                   & $705.14$ \\
                   &
                   & \quad$\pm 2.58$
                   & 
                   &    &
                   & \quad$\pm 2.81$
                   & \quad$\pm 0.98$
                   & \quad$\pm 0.82$
                   & \quad$\pm 0.98$
                   & \quad$\pm 0.009$
                   & \quad$\pm 4.05$
                   &  &  &  &  & \\
28000 & B\tablenotemark{e} 
                   & $65.50$
                   & $0.573$
                   & \nodata
                   & \nodata
                   & $40.56$
                   & $ 9.31$
                   & $19.85$
                   & $19.26$
                   & $3.450$
                   & $ 93.49$
                   & $27.53$
                   & $49.29$
                   & $35.33$
                   & $551.63$
                   & $664.20$ \\
                   &
                   & \quad$\pm 1.59$
                   & \quad$\pm 0.039$
                   &    &
                   & \quad$\pm 4.14$
                   & \quad$\pm 0.58$
                   & \quad$\pm 0.85$
                   & \quad$\pm 0.98$
                   & \quad$\pm 0.008$
                   & \quad$\pm 5.97$
                   &  &  &  &  & \\
\enddata
\tablenotetext{a}{The stellar radius derived from the models.}
\tablenotetext{b}{The stellar radius derived from the measured parallax and the assumed temperature.}
\tablenotetext{c}{The projected rotational velocity of the O-star derived from the models.  The observed
value is $95\pm 6$ km s$^{-1}$.}
\tablenotetext{d}{Model A assumes a circular orbit and synchronous rotation.}
\tablenotetext{e}{Model B assumes a circular orbit and nonsynchronous rotation.}
\tablenotetext{f}{Model C assumes an eccentric orbit and pseudo-synchronous rotation}
\tablenotetext{g}{Model D assumes an eccentric orbit and nonsynchronous rotation}

\end{deluxetable}

\clearpage

\pagestyle{plain}

\setcounter{figure}{0}
\renewcommand{\thefigure}{A\arabic{figure}}

\begin{figure}
\epsscale{0.7}
\plotone{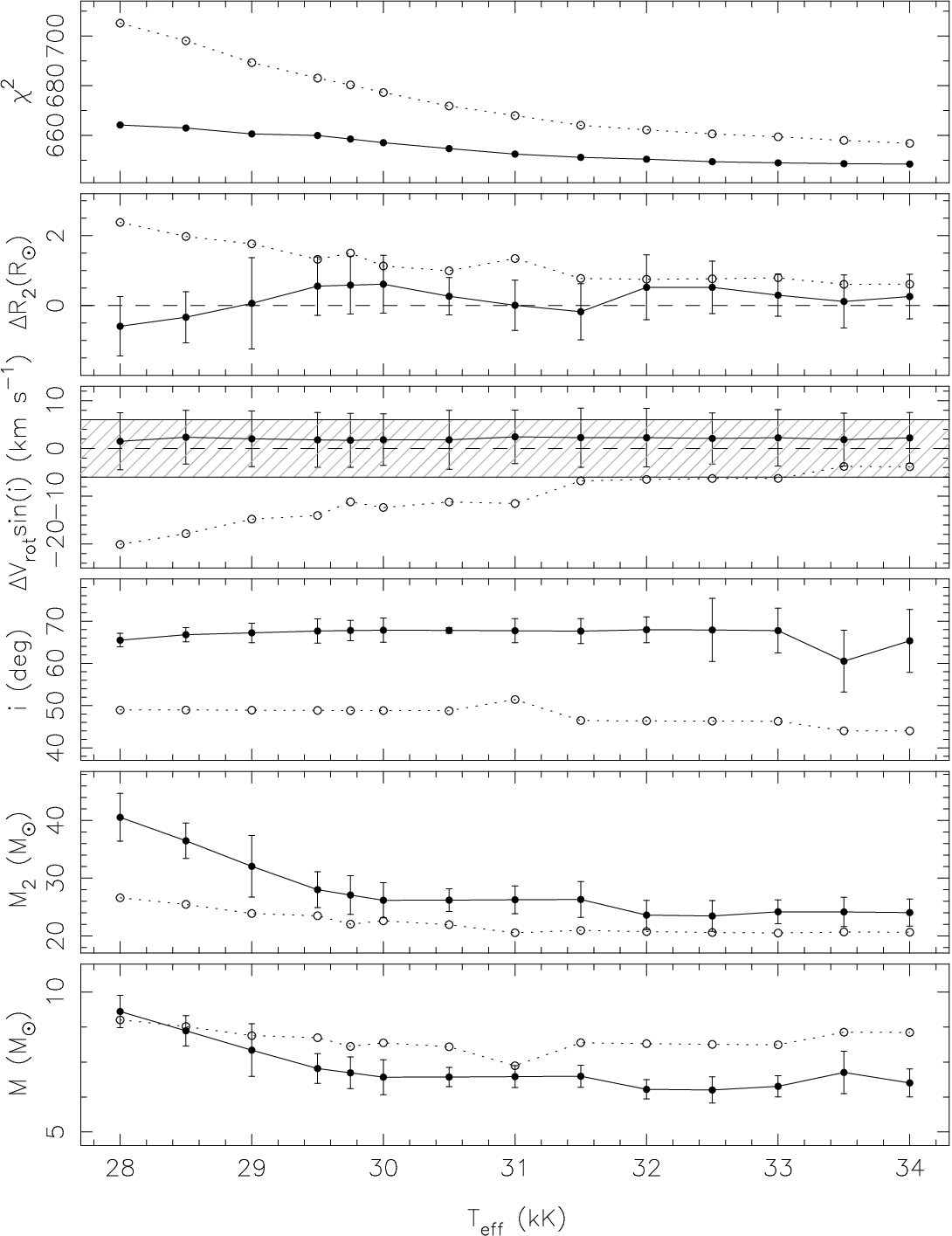}
\caption{Results of the dynamical modeling assuming a circular orbit.
Various quantities are plotted as a function of temperature
for the model assuming synchronous rotation
(open circles) and nonsynchronous rotation (filled circles).
From top to bottom the plots show
(a) the total $\chi^2$;
(b) the difference between $R_{\rm dist}$ and $R_2$ in solar radii;
(c) the difference between the observed projected rotational velocity
and the value derived from the models in km s$^{-1}$, where the hatched
region denotes the $1\sigma$ uncertainty on the measured value;
(d) the inclination in degrees;
(e) the mass of the O-star in solar masses;
and (f) the mass of the black hole in solar masses.  In general, for the
synchronous models the O-star
radius derived from the 
models is smaller than the radius derived from the distance, whereas
the derived projected rotational velocity is larger than observed.}
\label{fig1circ}
\end{figure}

\clearpage

\begin{figure}
\epsscale{0.8}
\plotone{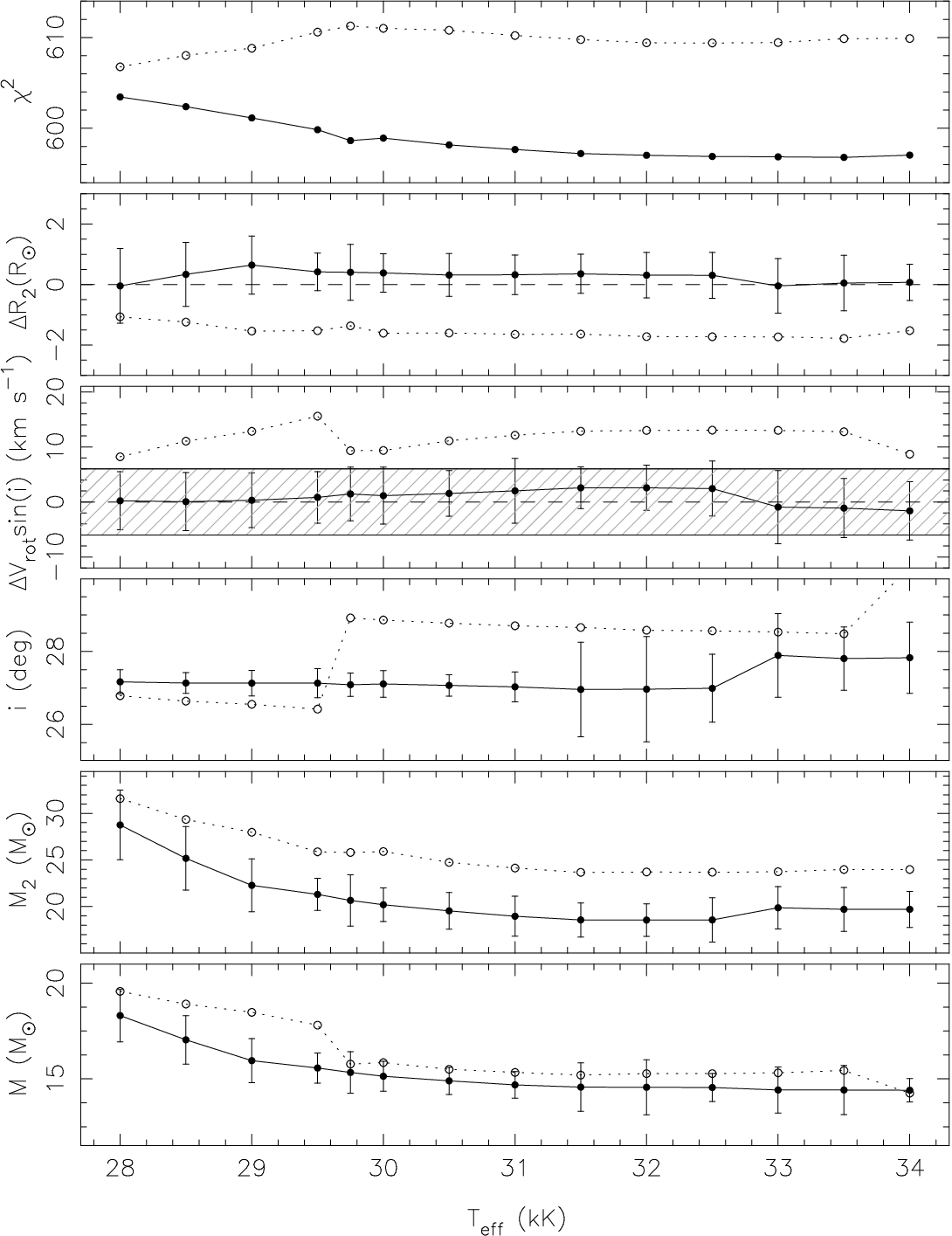}
\caption{Same as Figure \protect{\ref{fig1circ}}, but for the models
that assume an eccentric orbit.  In this case, the O-star radius
derived from the models is larger than the radius derived from the
distance, and the derived rotational velocity is smaller than
observed.}
\label{fig1ecc}
\end{figure}


\begin{thebibliography}{}

\bibitem[Abubekerov, et al.(2005)]{abu+2005}
{Abubekerov}, M. K.,
{Antokhina}, {\'E}. A., \& 
{Cherepashchuk}, A. M. 2005, 
Astron.\ Rep., 49, 801

\bibitem[Aufdenberg et al.(1998)]{auf+1998}
{Aufdenberg}, J. P.,
{Hauschildt}, P. H.,
{Shore}, S. N., \&
{Baron}, E. 1998,
\apj, 498, 837

\bibitem[Bolton(1975)]{bol+1975}
{Bolton}, C. T. 1975, 
\apj, 200, 269


\bibitem[Brocksopp et al.(1999a)]{bro+1999a}
{Brocksopp}, C.,
{Tarasov}, A.~E.,
{Lyuty},  V.~M., \&
{Roche}, P. 1999a,
\aap, 343, 861

\bibitem[Brocksopp et al.(1999b)]{bro+1999b}
{Brocksopp}, C.,
{Fender}, R.~P.,
{Larionov}, V.,
{Lyuty},  V.~M., 
{Tarasov}, A.~E.,
{Pooley}, G.~G.,
{Paciesas}, W.~S., \&
{Roche}, P. 1999b,
\mnras, 309, 1063

\bibitem[Caballero-Nieves et al.(2009)]{cab+2009}
{Caballero-Nieves}, S. M., et~al.\ 2009,
\apj, 701, 1895

\bibitem[Cadolle Bel et al.(2006)]{cad+2006}
Cadolle Bel, M., Sizun, P., Goldwurm, A., Rodriguez, J., Laurent, P.,
Zdziarski, A. A., Foschini, L., Goldoni, P., et al.\ 2006, \aap, 446, 591


\bibitem[Cardelli et al.(1989)]{car+1989}
{Cardelli}, J.~A.,
{Clayton}, G.~C., \&
{Mathis}, J.~S. 1989,
\apj, 345, 245

\bibitem[Eaton(2008)]{eat+2008}
{Eaton}, J. A. 2008,
\apj, 681, 562

\bibitem[Eggleton(1983)]{egg+1983}
{Eggleton}, P. P. 1983,
\apj, 268, 368

\bibitem[Frontera et al.(2001)]{fro+2001}
Frontera, F., Palazzi, E., Zdziarski, A. A., Haardt, F., Perola, G. C.,
Chiappetti, L., Cusumano, G., Dal Fiume, D., et al.\ 
2001, \apj, 546, 1027
        


\bibitem[Gies \& Bolton(1986)]{gie+1986}
{Gies}, D. R., \&
{Bolton}, C.~T. 1986,
\apj, 304, 371

\bibitem[Gierlinski et al.(1997)]{gie+1997}
Gierlinski, M., Zdziarski, A. A., Done, C., Johnson, W. N., Ebisawa, K.,
Ueda, Y., Haardt, F., \& Phlips, B. F. 1997, MNRAS, 288, 958


\bibitem[Gies et al.(2003)]{gie+2003}
{Gies}, D. R., et al.\ 2003, 
\apj, 583, 424 (2003).

\bibitem[Gou et al.(2011)]{gou+2010}
{Gou}, L. J., et al.\ 2011,
\apj, submitted

\bibitem[Guinan et al.(1979)]{gui+1979}
Guinan, E. F.,
{Dorren}, J.~D.,
{Siah}, M.~J., \&
{Koch}, R.~H.  1979, 
\apj, 229, 296

\bibitem[Herrero et al.(1995)]{her+1995}
{Herrero}, A.,
{Kudritzki}, R.~P.,
{Gabler},  R.,
{Vilchez}, J.~M., \&
{Gabler}, A. 1995,
\aap, 297, 556

\bibitem[Hutchings(1978)]{hut+1978}
{Hutchings}, J. B. 1978,
\apj, 226, 264


\bibitem[Kalogera \& Baym(1996)]{kal+1996}
{Kalogera}, V., \& {Baym}, G. 1996,
\apjl, 470, L61 


\bibitem[Karitskaya et al.(2005)]{kar+2005}
{Karitskaya}, E.~A., et~al., 2005
Astron.\ Astrophys.\ Trans., 24, 383

\bibitem[Lanz \& Hubeny(2003)]{lan+2003}
Lanz, T., \&  Hubeny, I. 2003,
\apjs, 146, 417

\bibitem[Lanz \& Hubeny(2007)]{lan+2007}
Lanz, T., \&  Hubeny, I. 2007,
\apjs, 169, 83


\bibitem[Ninkov et al.(1987)]{nin+1987}
{Ninkov}, Z.,
{Walker}, G.~A.~H., \&
{Yang}, S. 1987,
\apj, 321, 425 

\bibitem[Orosz \& Hauschildt(2000)]{oro+2000}
{Orosz}, J. A., \& Hauschildt, P. H. 2000,
\aap, 364, 265 

\bibitem[Orosz et al.(2002)]{oro+2002}
{Orosz}, J. A., et al.\ 2002,
\apj, 568, 845


\bibitem[Orosz et al.(2007)]{oro+2007}
{Orosz}, J. A., et al.\ 2007, 
\nat, 449, 872

\bibitem[Orosz et al.(2009)]{oro+2009}
{Orosz}, J. A., et al.\ 2009,
\apj, 697, 573

\bibitem[Paczy\'nski(1974)]{pac+1974}
Paczy\'nski, B. 1974, \aap, 34, 161

\bibitem[Reid et al.(2011)]{rei+2010}
{Reid}, M. J., et al.\ 2011,
\apj, in press

\bibitem[Remillard \& McClintock(2006)]{rem+2006}
Remillard, R. A., \& McClintock, J. E., 2006, ARAA, 44, 49


\bibitem[Shaposhnikov \& Titarchuk(2007)]{sha+2007}
{Shaposhnikov}, N., \& {Titarchuk}, L. 2007,
\apj, 663, 445

\bibitem[Skrutskie et al.(2006)]{skr+2006}
{Skrutskie}, M. F., et al.\ 2006,
\aj, 131, 1163 

\bibitem[Trimble et al.(1973)]{tri+1973}
{Trimble}, V.,
{Rose}, W.~K., \&
{Weber}, J. 1973,
\mnras, 162, 1P

\bibitem[Wilson \& Sofia(1976)]{wil+1976}
{Wilson}, R. E., \& {Sofia}, S. 1976,
\apj, 203, 182 


\bibitem[Wilson(1990)]{wil+1990}
Wilson, R. E. 1990,
\apj, 356, 613

\bibitem[Zhang et al.(1997)]{zha+1997}
Zhang, S. N., Cui, W., Harmon, B. A., Paciesas, W. S., Remillard, R. E., \&
van Paradijs, J. 1997, ApJ, 477, L95




\end{thebibliography}
\end{document}